\documentclass[fleqn,usenatbib]{mnras}

\usepackage{amsmath}
\usepackage{amssymb}
\usepackage{graphicx}
\usepackage{epsf}
\usepackage{subfigure}
\usepackage{color}
\usepackage{threeparttable}
\usepackage{dcolumn}
\usepackage{comment}
\usepackage{epsfig}
\usepackage{xspace}
\usepackage{multirow}
\usepackage{makecell}
\usepackage{appendix}
\usepackage{booktabs}

\newcommand{\orc}{\includegraphics[height=\fontcharht\font`A]{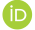}}

\usepackage{array}
\date{\today}

\DeclareGraphicsExtensions{.jpg,.pdf,.png,.eps,.ps}

\newcommand*{\TT}{\ensuremath{TT}}
\newcommand*{\EE}{\ensuremath{E\!E}}
\newcommand*{\TE}{\ensuremath{T\!E}}
\newcommand*{\lcdm}{$\Lambda$CDM}
\newcommand*{\planck}{\textit{Planck}}

\newcommand*{\Nreal}{\ensuremath{N_{\mathrm{real}}}}
\newcommand*{\Nsim}{\ensuremath{N_{\mathrm{sim}}}}

\newcommand{\RNum}[1]{\uppercase\expandafter{\romannumeral #1\relax}}

\newcommand*{\sqdeg}{\ensuremath{{\rm deg}^2}}

\defcitealias{dutcher21}{D21}
\defcitealias{riess20}{R20}

\definecolor{amber}{rgb}{1.0, 0.49, 0.0}

\newcommand{\skipt}[1]{}
\begin{document}

\title[Parameter-level Performance of Covariance Conditioning]{The Parameter-Level Performance of Covariance Matrix Conditioning in Cosmic Microwave Background Data Analyses}

\author[L.~Balkenhol and C.~L.~Reichardt]{
L.~Balkenhol$\,{\href{https://orcid.org/0000-0001-6899-1873}{\orc}}$\thanks{Corresponding author: \href{mailto:lbalkenhol@student.unimelb.edu.au}{lbalkenhol@student.unimelb.edu.au}}
and C.~L.~Reichardt$\,{\href{https://orcid.org/0000-0003-2226-9169}{\orc}}$
\\
School of Physics, University of Melbourne, Parkville, VIC 3010, Australia}


\maketitle

\begin{abstract}
Empirical estimates of the band power covariance matrix are commonly used in cosmic microwave background (CMB) power spectrum analyses.
While this approach easily captures correlations in the data, noise in the resulting covariance estimate can systematically bias the parameter fitting. 
Conditioning the estimated covariance matrix, by applying prior information on the shape of the eigenvectors, can reduce these biases and ensure the recovery of robust parameter constraints.
In this work, we use simulations to benchmark the performance of four different conditioning schemes, motivated by contemporary CMB analyses.
The simulated surveys measure the \TT{}, \TE{}, and \EE{} power spectra over the angular multipole range $300 \le \ell \le 3500$ in $\Delta \ell = 50$ wide bins, for temperature map-noise levels of $10, 6.4$ and $2\,\mu$K-arcmin.
We divide the survey data into $\Nreal = 30, 50,$ or 100 uniform subsets.
We show the results of different conditioning schemes on the errors in the covariance estimate, and how these uncertainties on the covariance matrix propagate to the best-fit parameters and parameter uncertainties. 
The most significant effect we find is an additional scatter in the best-fit point, beyond what is expected from the data likelihood. 
For a minimal conditioning strategy, $\Nreal = 30$, and a temperature map-noise level of 10$\,\mu$K-arcmin, we find the uncertainty on the recovered best fit parameter to be $\times 1.3$ larger than the apparent posterior width from the likelihood ($\times 1.2$ larger than the uncertainty when the true covariance is used). 
Stronger priors on the covariance matrix reduce the misestimation of parameter uncertainties to $<1\%$.
As expected, empirical estimates perform better with higher $\Nreal$, ameliorating the adverse effects on parameter constraints.
\end{abstract}


\section{Introduction}
\label{sec:intro}

Current cosmological surveys provide new insight on inflation, dark matter, dark energy, and neutrino properties \citep{DE2013, Abazajian2015, planck2018, dutcher21, aiola20}.
Cosmic microwave background (CMB) and large-scale structure (LSS) observations complement one another by providing snapshots of the universe at different times.
The interplay of these two probes allows us to constrain structure formation and modified gravity models \citep{Omori2019}.

A crucial ingredient in many data analyses is the covariance matrix, which encodes the uncertainties on the measured data points.
An accurate covariance matrix is essential for producing reliable model constraints; errors in the covariance matrix directly propagate into skewed cosmological parameters \citep{Hartlap2007, Hamimeche2009, Taylor2013, Dodelson2013}.

Since analytic calculations of the covariance matrix may struggle to capture the full complexity of the instrument, estimators are frequently used on simulated or real data to generate a robust estimate of the covariance matrix. 
While one can use simulations to produce a covariance estimate to any desired precision, this requires a good understanding of the noise properties of the experiment, and, perhaps more importantly, can be very computationally expensive. 
\citet{Dodelson2013} note that modern surveys underestimate their parameter uncertainties by $\sim$\,5-10\%, due to only running $\mathcal{O}(100)$ simulations (e.g., \citet{Louis2017, henning2018, Planck2018Spectra}).
Shrinkage estimators \citep{Hamimeche2009}, the \textsc{CARPool} algorithm \citep{chartier21}, data compression via \textsc{MOPED} \citep{heavens17}, emulators \citep{Schneider2008, Morrison2013}, and large-scale mode-resampling methods \citep{Schneider2011} can ameliorate the increasing computational cost of simulations by reducing the number of realisations required to achieve adequate precision.

Alternatively, the measured data itself can be used to estimate the covariance matrix \citep[e.g.,][]{lueker2010, Das2014, Crites2015, henning18, dutcher21}.
This approach does not make any assumptions about the noise profile of the experiment and can avoid or reduce the required number of computationally expensive simulations. 
Empirical estimators require that the data is divided into uncorrelated, uniform subsets, i.e. independent realisations. 
The specific details of how the survey was conducted generally impose an upper limit on how many such subsets can be created, which can limit the accuracy of the covariance matrix estimate. 

Conditioning can be applied to more accurately estimate the covariance matrix obtained from any of the above methods \citep[e.g.,][]{lueker2010, Ade2014, Das2014, bkplanck2015, Crites2015, Louis2017, henning2018, choi20,  dutcher21}.
Conditioning schemes introduce prior knowledge of the correlation structure of the covariance matrix or of the values of specific entries, in order to reduce the overall uncertainty on the covariance estimate.
The details of the conditioning approach vary drastically between experiments, ranging from no conditioning to strong priors.
The approach chosen for a given analysis is often not justified in detail, and the effects on the final parameter constraints are rarely discussed. 

This work fills this gap.
Focussing on CMB data analysis, we investigate the impact of conditioning at the covariance- and the parameter-level.
This is done using a suite of simulations, mimicking access to $\Nreal = 30, 50,$ or 100 data realisations and temperature-map noise levels of $\sigma_{\rm map} = 10, 6.4$ and $2\,\mu$K-arcmin. 
We assume the \TT{}, \TE{} and \EE{} power spectra are measured over the angular multipole range $300 \le \ell \le 3500$ in bins with$\Delta \ell = 50$.
We report results for four different conditioning strategies, paying special attention to signs of bias and uncertainty misestimation in the parameter constraints.
We also look briefly at how these results shift when changing the combination of temperature and polarisation spectra analysed, and the width of band power bins.

This work is structured as follows.
In Section \S\ref{sec:theory}, we review the necessary background information on the covariance matrix, how to estimate it empirically, and define the four conditioning schemes we benchmark later on.
In Section \S\ref{sec:sims}, we describe the simulation framework used.
We analyse our results in Section \S\ref{sec:results}, first in covariance space, then at the parameter-level, before concluding in Section \S\ref{sec:conclusion}.

\section{Band Power Covariance}
\label{sec:theory}

\subsection{Empirical Estimation}
\label{subsec:bootstrapping}

The band power covariance matrix, $\mathcal{C}$, describes the uncertainty on band power measurements and the correlations between them. It is defined as
\begin{equation}
\mathcal{C}_{bb'} = \left\langle \left( C_b - \overline{C}_b \right) \left( C_{b'} - \overline{C}_{b'} \right) \right\rangle,
\end{equation}
where $C_b$ is the binned power spectrum and $\overline{C}_b$ the average thereof \citep[e.g.,][]{master, tristram05}.
\citet{tristram05} show that under certain assumptions we can model the covariance matrix as
\begin{equation} \label{eq:cross_cov_simple}
\mathcal{C}^{AB,CD}_{bb'} \simeq \frac{1}{\nu_b}\left( C^{AC}_b C^{BD}_b + C^{AD}_b C^{BC}_b \right) \delta_{bb'},
\end{equation}
where $\{A,B,C,D\} \in \{T, E, B\}$ are any temperature or polarisation field of the CMB, and $\nu_b$ is the effective number of modes in a given band power bin.

However, modelling the experimental noise sufficiently accurately to use these theoretical prescriptions can be difficult.
Eqn. \ref{eq:cross_cov_simple} also assumes the mode-coupling matrix can be treated as diagonal, which is rarely satisfied given filtering choices and the requirement to mask bright radio galaxies and the Milky Way. 
The mode-coupling reduces the accuracy of straight-forward analytic methods to $\mathcal{O}(10\%)$ \citep{Hamimeche2009}.

Alternatively, the covariance matrix can be estimated empirically.
This approach relies on forming a uniform set of $\Nreal$ realisations of the experimental data, by dividing it into uniform subsets.
These realisations form the basis of estimating the noise and signal-noise interaction term of the covariance matrix, which we refer to together as the noise-variance contribution, $\mathcal{C}^{N}_{b b'}$.
Adding the signal term, i.e. the sample-variance, $\mathcal{C}^{S}_{b b'}$, to the noise-variance gives the full band power covariance matrix: $\mathcal{C}_{b b'} = \mathcal{C}^{S}_{b b'} + \mathcal{C}^{N}_{b b'}$.

By calculating the $\Nreal(\Nreal-1)/2$ unique cross spectra of the data realisations, $\hat{C}^{\alpha \beta}_\ell, \alpha \neq \beta$, where $\alpha$ and $\beta$ index the realisation, we can use the observed scatter of the distribution of band powers to estimate part of the covariance matrix.
As shown by \citet{lueker10}, the correct noise-variance contribution to the covariance, is provided by the estimator
\begin{equation} \label{eq:cov_full_est}
\begin{aligned}
\hat{\mathcal{C}}^{N}_{b b'} = \frac{2 f(\Nreal)}{\Nreal^4} \sum_\lambda \sum_{\alpha \neq \lambda} &\left[ \Delta C^{\lambda \alpha}_b \Delta C^{\lambda \alpha}_{b'} \right. \\
&\left. + 2 \left( \sum_{\beta \neq \lambda, \alpha} \Delta C^{\lambda \alpha}_{b} \Delta C^{\lambda \beta}_{b'} \right) \right],
\end{aligned}
\end{equation}
where $\Delta C_{b}$ are the mean-subtracted binned power spectra and $f(\Nreal)$ is a correction for the finite number of realisations that approaches unity for large $\Nreal$.
In this limit, $\langle \hat{\mathcal{C}}^{N}_{b b'} \rangle = \mathcal{C}^{N}_{b b'}$.
The sample-variance contribution can be obtained from the signal-only simulations typically used for the transfer-function calculation; since these simulations are uncorrelated, the variance of band powers provides an unbiased estimate of $\mathcal{C}^{S}_{b b'}$.
Therefore, by adding the noise-variance estimator to the variance of the signal-only simulation spectra, the band power covariance matrix can be estimated.

Empirical estimators provide a noisy estimate of the band power covariance matrix, where the number of realisations $\Nreal$ sets the precision.
\citet{lueker2010} show that each element has a statistical variance of
\begin{equation} \label{eq:cov_error}
\left\langle \left( \mathcal{C}_{b b'} - \left\langle \mathcal{C}_{b b'} \right\rangle \right)^2 \right\rangle = \frac{\mathcal{C}_{b b'}^2 + \mathcal{C}_{b b} \mathcal{C}_{b' b'}  }{\Nreal}.
\end{equation}
Errors in the covariance matrix estimate propagate through to cosmological parameters, widening the posterior distributions and biasing the best-fit point \citep{Hartlap2007, Hamimeche2009, Taylor2013, Dodelson2013}.
Increasing $\Nreal$ reduces the errors, and thus the magnitude of these effects, although the observation strategy often limits the possible number of data splits for real data. 
For simulations, the computing cost limits the number of realisations available.

\subsection{Conditioning}
\label{subsec:conditioning}

Conditioning the covariance matrix estimate can reduce the errors for a fixed $\Nreal$. 
This step takes place before parameter estimation and uses prior knowledge to constrain noisy elements of the covariance matrix.
The details of conditioning schemes vary drastically between analyses.
For example, \citet{Ade2014}, \citet{Das2014} and \citet{Louis2017} drop the off-diagonal elements of all blocks, i.e. ignore mode-coupling.
\citet{choi20} allow for mode-coupling to the nearest angular multipole bin, but explicitly zero the effects of filtering and masking further than that.
Similarly, \citet{Ade2018} keep correlations in the first off-diagonal bins, but further reduce the noise of the covariance estimate by explicitly ignoring contributions with an expectation value of zero.
\citet{Crites2015} and \citet{henning2018} both average band-diagonal elements in the auto-correlation matrices to address masking effects.
A direct manipulation of on-diagonal elements is performed by \citet{dutcher21} and \citet{reichardt20} (as well its preceding analyses targeting secondary anisotropies).

To navigate this broad landscape we define four different, increasingly strict conditioning strategies, which we will benchmark in this work. These act on the complete covariance matrix, i.e. the sum of the noise-variance and sample-variance estimators.
\begin{itemize}
\item \emph{No conditioning:} The final covariance is given by the noise-variance plus sample-variance estimators.
\item \emph{Diagonal conditioning:} We calculate the diagonal elements of all covariance blocks as before, but set all off-diagonal elements to zero.
\item \emph{Moderate conditioning:} Diagonal elements are calculated as before.
Off-diagonal elements of the auto-covariance blocks are band-averaged in the correlation matrix for bins within $\Delta \ell \leq 100$ of the diagonal, but set to zero for $\Delta \ell > 100$.
All off-diagonal elements of off-diagonal covariance blocks are set to zero.
\item \emph{Deep conditioning:} An estimate of the effective number of modes and the noise spectrum is extracted from a combination of real data and transfer-function simulations.
Using this information, diagonal elements are replaced by their theoretical expectation values.
Second order polynomials are fitted to band-diagonal elements in the correlation matrices (for details see \citet{dutcher21}).
\end{itemize}

\section{Simulation Framework}
\label{sec:sims}

We run simulations to gauge to what extend the conditioning schemes outlined in \S\ref{subsec:conditioning} protect against the adverse effects of a noisy covariance matrix estimate.
To reduce the computational cost, we mimic the observation of a modest  $\sim 290\,\sqdeg$ field, comprised of $512 \times 512$ square pixels of $2\,\mathrm{arcmin}$ width, using the flat-sky approximation. 
The underlying CMB signal is based on a power spectrum calculated by \textsc{camb} \citep{lewis00}\footnote{https://camb.info/}, with cosmological parameters set to the latest \planck{} results \citep{planck18-6}.
We add white and 1/f noise at the map level, which follows the power spectrum
\begin{equation}
N^{X}_{\ell} = N^{X}_{\rm white} \left[ 1 + \left( \frac{\ell}{\ell_{\rm knee}} \right)^{\alpha^{X}} \right],
\end{equation}
where $X \in \{T, E\}$, $\alpha^T = -3.5$, $\ell^T_{\rm knee} = 1000$, $\alpha^E = -1.4$, $\ell^E_{\rm knee} = 700$, and $N^{X}_{\rm white}$ is the white noise level, which is in turn set by the map-noise level, $\sigma_{\rm map}$.
We assume that the white noise power in polarisation is twice that of temperature.
This noise curve is chosen to approximate the anticipated noise characteristics of the Simons Observatory (SO) large aperture telescope (c.f. Equation 1 in \citet{simonsobservatorycollab19}).
We simulate having access to $\Nreal$ maps of the same CMB signal with independent noise realisations added to each one.
We proceed to measure the \TT{}, \TE{}, and \EE{} power spectra across the angular multipole range $300 \leq \ell < 3500$ and bin them into band powers of width $\Delta\ell = 50$.
The simulated power spectrum measurements are complemented by $\Nsim=300$ signal-only simulations, which mimic the typical approach to calculating the transfer-function.

We run nine sets of simulations, for $\Nreal = 30$, 50 or 100, and $\sigma_{\rm map} = 10$, 6.4, or 2\,$\mu$K-arcmin.
The number of realisations are chosen to span the range seen in contemporary analyses, without inflating computation cost \citep[e.g.,][]{dutcher21, henning18, choi20, louis17, planck18-5}.
The map noise levels are chosen to match the SO baseline configuration, the SO goal, and the projected depth of the complete SPT-3G survey \citep{simonsobservatorycollab19, bender18}.
We run $1000, 500, 100$ simulations for $\Nreal = 30, 50, 100$, respectively.

For each simulation we estimate the band power covariance matrix as detailed in \S\ref{subsec:bootstrapping} and condition copies of it as described in \S\ref{subsec:conditioning}, i.e. each simulation is associated with four differently conditioned versions of the same covariance matrix estimate.
Furthermore, we calculate the average of all unconditioned covariance matrices for a given $\Nreal, \sigma_{\rm map}$ combination, which we refer to as the fiducial covariance matrix.
We also calculate a theoretical covariance void of mode-coupling using the prescriptions in Equation \ref{eq:cross_cov_simple}.

We note that the sky area chosen here is much less than the coverage of current and next-generation ground-based CMB surveys.
However, the sky fraction only serves to scale the overall amplitude of the covariance matrix.
Ultimately, we are interested in a comparative study of conditioning schemes, not their absolute performance.
Therefore the shape of the covariance matrix and the noise on the covariance estimate, as set by $\sigma_{\rm map}$ and $\Nreal$, respectively, are of greater interest here.

\section{Results}
\label{sec:results}

\subsection{Comparison in Covariance Space}
\label{subsec:cov_comparison}

\begin{figure*}
\includegraphics[width=17.2cm]{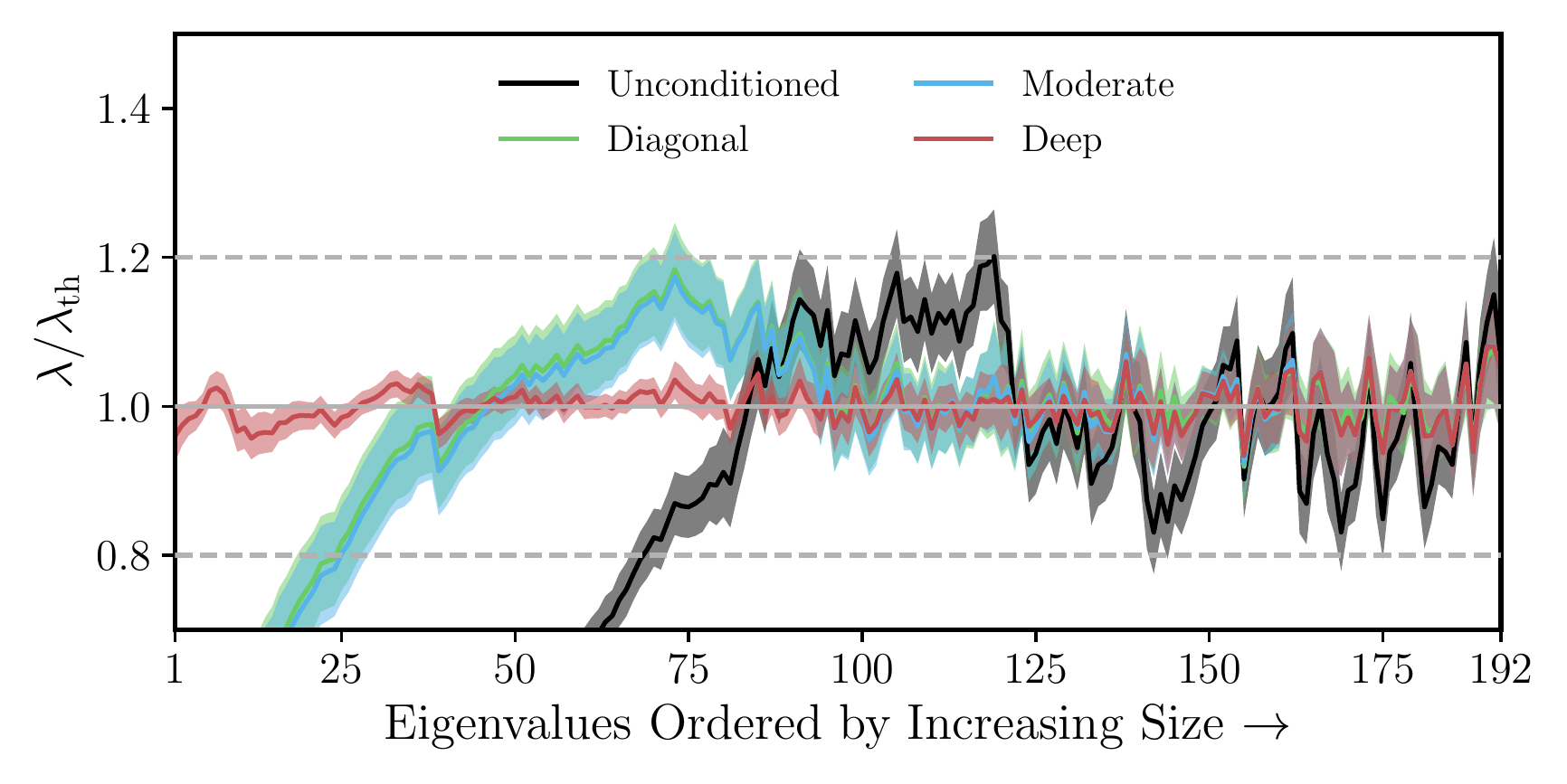}
\centering
\caption{
Comparison of the eigenvalues of covariance matrices for different conditioning strategies: unconditioned (black), diagonal (green), moderate (blue), deep (red). We plot the ratio of the average of the absolute eigenvalues across all $\Nreal=30, \sigma_{\rm map} = 10\,\mu$K-arcmin simulations in increasing order, to the eigenvalues of the theoretical covariance matrix. The shaded regions correspond to the standard deviation around the mean.
The eigenvalues of unconditioned and mildly conditioned covariances recover the largest third and half of eigenvalues well, respectively - beyond that the eigenvalues drop off.
The dotted grey lines indicate our threshold for acceptable deviations from the fiducial case, $\leq20\%$.
For the unconditioned covariances $39\%$ of values lie outside this range, whereas $14\%$ of the eigenvalues of diagonal and moderately conditioned matrices show large deviations from the expectation.
Deep conditioning recovers the entire range of eigenvalues.
}
\label{fig:eig}
\end{figure*}

First, we look at how well the true covariance matrix can be estimated, for a fixed number of realisations, under each conditioning scheme.
We assess this by looking at the eigenvalues of differently conditioned covariance matrices for the $\Nreal=30$, $\sigma_{\rm map} = 10\,\mu$K-arcmin simulations in Figure \ref{fig:eig}.

We see that the largest $\sim 70$ eigenvalues are recovered well by all conditioning schemes, though the unconditioned matrices show the largest fluctuations around the theoretical expectation values.
We consider fluctuations $>20\%$ away from the fiducial values as failures to retrieve the correct eigenvalues.
For the unconditioned matrices, $61\%$ of eigenvalues are within this threshold.
The performance of the other conditioning schemes is comparable to one another for the largest $\sim 70$ eigenvalues, though for smaller eigenvalues they begin to deviate; $14\%$ of the eigenvalues of the diagonal and moderately conditioned covariance deviate by more than $20\%$ from the fiducial case.
Deeply conditioned matrices recover all eigenvalues, with no statistically significant deviation from the theoretical covariance.

Noisy estimates of the covariance matrix can lead to negative eigenvalues, which remain a problem for most of the conditioning strategies. 
Deep conditioning is the exception, yielding positive-definite covariance matrices in all cases tested.
As the breakdown in positive definiteness is related to the magnitude of the noise on the covariance estimate, increasing the number of realisations generally improves the situation  (c.f. Equation \ref{eq:cov_error}).
For example, for the $\sigma_{\rm map} = 10\,\mu$K-arcmin simulations, $24\%$, $19\%$, and $12\%$ of the eigenvalues of unconditioned matrices are negative for $\Nreal = 30, 50,$ and $100$, respectively.
Diagonal and moderate conditioning perform better: at $\Nreal = 30$ only $0.4\%$ and $0.3\%$ of eigenvalues are negative, which reduces to $0.03\%$ and $0.02\%$ for $\Nreal = 50$, respectively.
For $\Nreal = 100$, diagonal and moderate conditioning match the performance of deep conditioning and we register no negative eigenvalues.

This work varies the number of realisations used in the empirical covariance estimate for the noise variance, but uses a fixed higher number of realisations to estimate the sample variance. 
As a result, the number of negative eigenvalues decreases in these tests as the angular multipole range dominated by sample variance increases (i.e.~as the map-noise level decreases). 
The sample-variance contribution to the covariance is always comparatively well-constrained with $300$ realisations. 
For example, the average fraction of negative eigenvalues for unconditioned covariance matrices reduces from $24\%$ for $\sigma_{\rm map} = 10\,\mu$K-arcmin, to $20\%$ and $14\%$ for $\sigma_{\rm map} = 6.4\,\mu$K-arcmin and $\sigma_{\rm map} = 2\,\mu$K-arcmin, respectively, for the $\Nreal=30$ simulations.

One can also improve the performance of empirical covariance estimators by reducing the number of entries in the covariance. 
In this context, the number can be reduced by using a coarser angular multipole binning for the power spectrum measurement. 
For the $\Nreal=30$, $\sigma_{\rm map} = 10\,\mu$K-arcmin simulations, the temperature and E-mode power spectrum is signal-dominated for $\ell < 3300$ and $\ell < 1750$, respectively.
Doubling the bin size to $\Delta\ell=100$ decreases the fraction of negative eigenvalues in the unconditioned covariances from $24\%$ to $17\%$.
Similarly, for diagonal and moderate conditioning the fraction of of negative eigenvalues is reduced to $0.02\%$ and $0.01\%$, respectively.
The performance of deep conditioning does not change.
While such rebinning can help reduce the noise of the covariance matrix, a coarser binning of the power spectrum can result in wider parameter constraints.

Given the large fraction of negative eigenvalues of the unconditioned matrices, we exclude this scheme from further analysis.

\subsection{Impact on Cosmological Parameters}
\label{subsec:parameter_comparison}

While errors on the covariance are easy to measure, the real statistics of interest are the effects on the derived parameter constraints during model fitting.
Our analysis focuses on the following key questions: is the parameter uncertainty over- or under-estimated? Is the best-fit point biased? Are the derived constraints suboptimal, i.e. do the results improve appreciably if the covariance estimate is improved?
Before pursuing these questions and reporting our results, we briefly describe our analysis methodology below.

We benchmark the effects on parameter fitting under the assumption of the \lcdm{} model, using model power spectra calculated by \textsc{camb} \citep{CAMB}. 
The six model parameters are: the amplitude of primordial density perturbations, $A_s$, the tilt of the power spectrum, $n_s$, defined at a pivot scale of $0.05\,\mathrm{Mpc^{-1}}$; the density of cold dark matter $\Omega_c h^2$; the baryon density $\Omega_b h^2$; the optical depth to reionisation, $\tau$; and $\theta_{MC}$, the approximation to ratio of the sound horizon to the angular diameter distance used by \textsc{camb}.

To reduce the computational cost of fitting cosmological parameters on $\mathcal{O}(5000)$ simulations, we primarly focus on a single parameter: $\theta_{MC}$, the approximation to the ratio of the sound horizon to the angular diameter distance.
We fix other cosmological parameters to the best-fit value of the signal-only spectrum obtained from combining all transfer-function simulations across the nine sets of simulations.
We explore the range $1.046912 \geq \theta_{MC} \geq 1.034112$ around this fiducial $\theta_{MC}$ value, although in practice we filter out any likelihoods that peak close to the edge of the range - outside of the $\sim 7.5\sigma$ region of the least constraining simulation.

We investigate the statistics in the full 6-dimensional parameter space for a subset of 100 of the $\Nreal = 30$, $\sigma_{\rm map} = 10\,\mu$K-arcmin simulations.
We use Cobaya \citep{Torrado2020} to find the minimum of the likelihood for these realisations.
Since the simulated surveys lack access to information from large angular scales, we fix the optical depth to reionisation $\tau$ to the input value.
Using the sampler of \citet{Lewis2002}, we run a full Markov Chain Monte Carlo for a single realisation using the fiducial band power covariance to obtain a fiducial parameter covariance matrix.

We calculate the joint likelihood of the \TT{}, \TE{}, and \EE{} band powers for each simulation using a Gaussian approximation:
\begin{equation} \label{eq:like}
\log{\mathcal{L}} = -0.5\,\Delta C_{b}\,\mathcal{C}_{bb'}^{-1}\,\Delta C_{b'},
\end{equation}
where $ \Delta C_b$ is the difference of the simulation and model band powers and $\mathcal{C}_{b b'}$ is the covariance matrix.
This procedure is carried out for each realisation using the diagonally-, moderately-, and deeply-conditioned covariance matrices, as well as the fiducial covariance.

We set any negative eigenvalues of the covariance matrices to a large, positive number, six orders of magnitude larger than maximum eigenvalue, in order to guarantee positive-definiteness for parameter estimation.
We analyse the likelihood curves of all simulations, identifying the peak of the likelihood as the best-fit point and obtaining the uncertainty of the parameter constraint from the curvature of the likelihood around this point.
We filter the results for outliers as follows.
First, we reject a realisation if the best-fit value of any conditioned matrix peaks towards the edge of the parameter range as mentioned before.
Furthermore, we perform sigma-clipping ($5\,\sigma$, five iterations), based on the best-fit values and parameter uncertainty.
For all simulations and analysis cases $>$\,96\% of realisations survive these cuts and we therefore do not expect this filtering to bias our results significantly.

We compare parameter uncertainties first and investigate the bias afterwards. 
We highlight key results in detail below, and point the reader to Tables \ref{tab:unc_bias} and \ref{tab:fields} in the Appendix for the full results.

\subsubsection{Uncertainty}
\label{subsubsec:unc}

\begin{figure*}
  \centering
  \begin{subfigure}{}
    \includegraphics[width=3.459in]{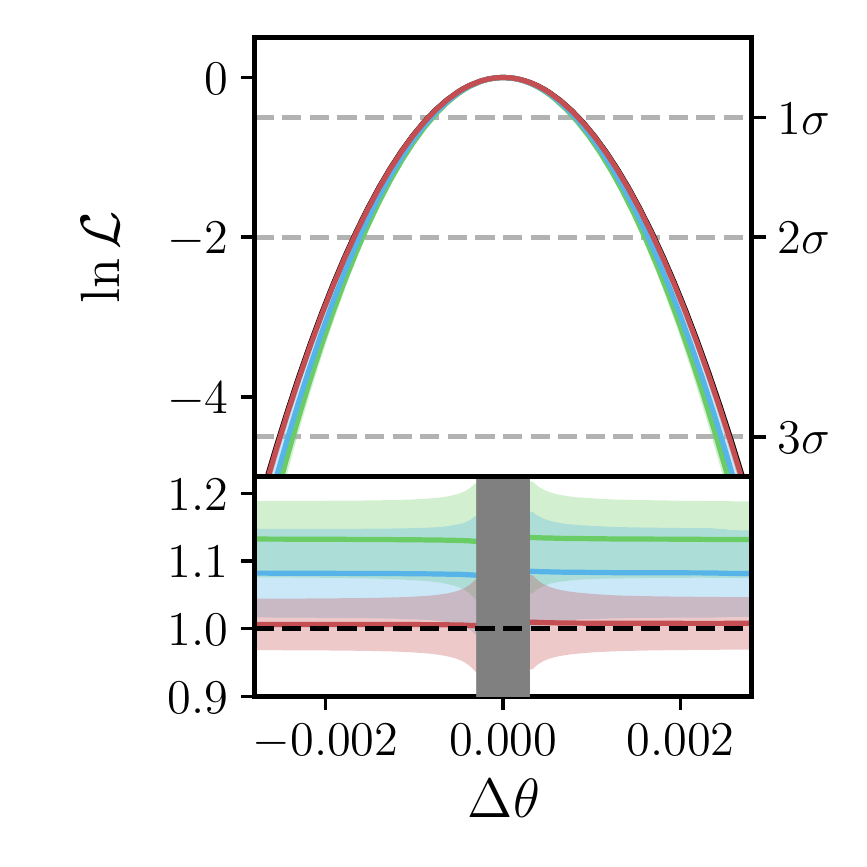}
  \end{subfigure}\hfill
  \begin{subfigure}{}
    \includegraphics[width=3.459in]{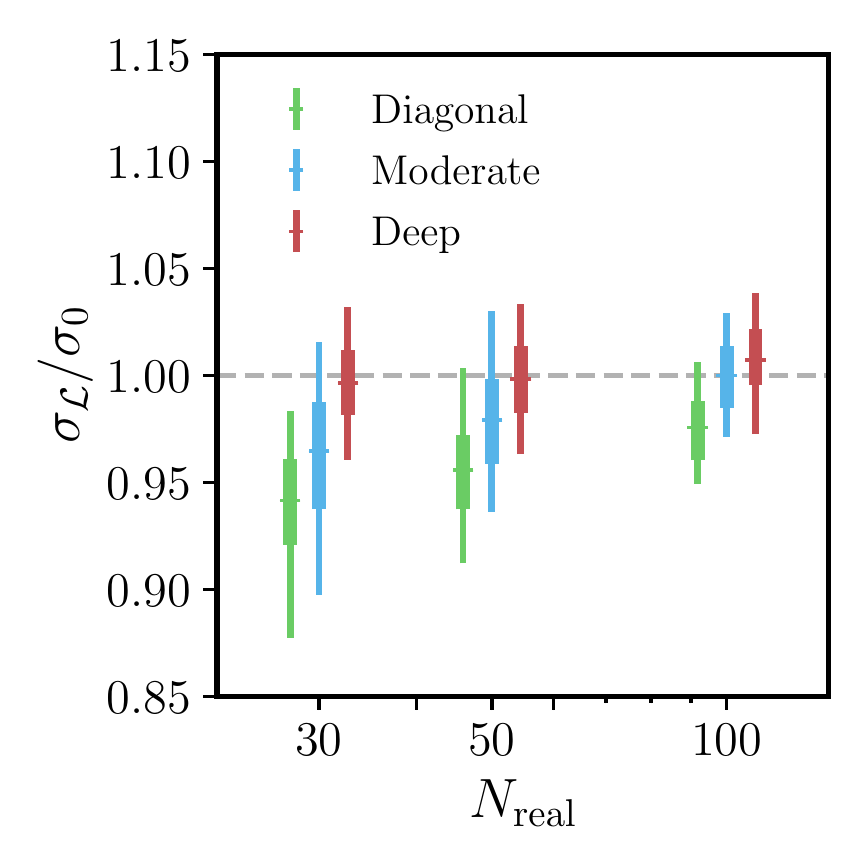}
  \end{subfigure}
  \caption{  \label{fig:logl}
\emph{Left panel:} We show stacked likelihood curves for $\Nreal = 30$, $\sigma_{\rm map} = 10\,\mu$K-arcmin simulations for diagonal (green), moderate (blue), and deep conditioning (red) and the fiducial covariance (black) in the top panel.
The shaded region around the lines corresponds to the standard deviation across simulations.
The curvature of the stacked likelihoods from diagonal, moderate, and deep conditioning deviate by 13\%, 8\%, 1\% from the fiducial covariance, respectively, which translates into percent-level differences in the uncertainty on $\theta_{MC}$.
The bottom panel shows the ratio of the different conditioning schemes to the stacked likelihood of the fiducial covariance.
We mask the region $|\Delta\theta| \leq 0.0003$ in dark grey, because of the numerical artefact in dividing by zero.\\
\emph{Right panel:}
We plot the median of the ratio of $\theta_{MC}$ uncertainty inferred from the data likelihood, $\sigma_{\mathcal{L}}$, from diagonal (green), moderate (blue), and deep conditioning (red), to the fiducial covariance matrix, $\sigma_0$, across all $\Nreal, \sigma_{\rm map} = 10\,\mu$K-arcmin simulations.
The wide and thin markers indicate the $68\%$ and $95\%$ interval, respectively.
While shallow conditioning schemes tend to underestimate the parameter uncertainty, increasing the number of independent realisations $\Nreal$ ameliorates this issue.
}
\end{figure*}

We inspect the uncertainty on parameter constraints for different conditioning schemes for the $\Nreal = 30$, $\sigma_{\rm map} = 10\,\mu$K-arcmin simulations and compare them to the fiducial covariance.
Diagonal covariance matrices underestimate the parameter uncertainty by $5.8^{+2.1}_{-1.9}\%$.
This underestimate is in-part due to zeroing the off-diagonal entries in the covariance matrix (which are non-zero due to mode-coupling), but also due to the excessive noise on the main diagonals of the covariance.
The latter point is confirmed by the performance of the moderately conditioned covariance matrices, which can recover the off-diagonal entries but have the same noisy estimate of the diagonals. 
The moderately conditioned case underestimates the uncertainty by $3.5^{+2.7}_{-2.3}\%$.
The deep conditioning scheme, which also reduces the uncertainty on the diagonal entries, recovers the fiducial error on $\theta_{MC}$ to within $0.3 \pm 1.5\%$. 
These differences between the conditioning schemes are illustrated in the left panel of Figure \ref{fig:logl}. 
The uncertainty on $\theta_{MC}$ is artificially changed by the uncertainty in the covariance matrix for the diagonal and moderate conditioning cases, while the likelihood for deep conditioning closely traces the fiducial likelihood curve. 

Increasing the number of realisations used in the covariance estimation will reduce the uncertainties (for a fixed number of bandpowers), and decrease their impact on the parameter uncertainties. 
The right panel of Figure \ref{fig:logl} shows the trend with increasing $\Nreal$ for $\sigma_{\rm map} = 10\,\mu$K-arcmin simulations.
As expected, the uncertainties on $\theta_{MC}$ converge to the fiducial uncertainty for all conditioning schemes as the number of realisations, $\Nreal$, increases. 
At $\Nreal = 100$, moderate conditioning recovers the fiducial uncertainty well, although diagonal conditioning still underestimates the uncertainty by $2.4^{+1.5}_{-1.2}\%$ due to zeroing the off-diagonal entries.

\begin{figure*}
\includegraphics[width=17.2cm]{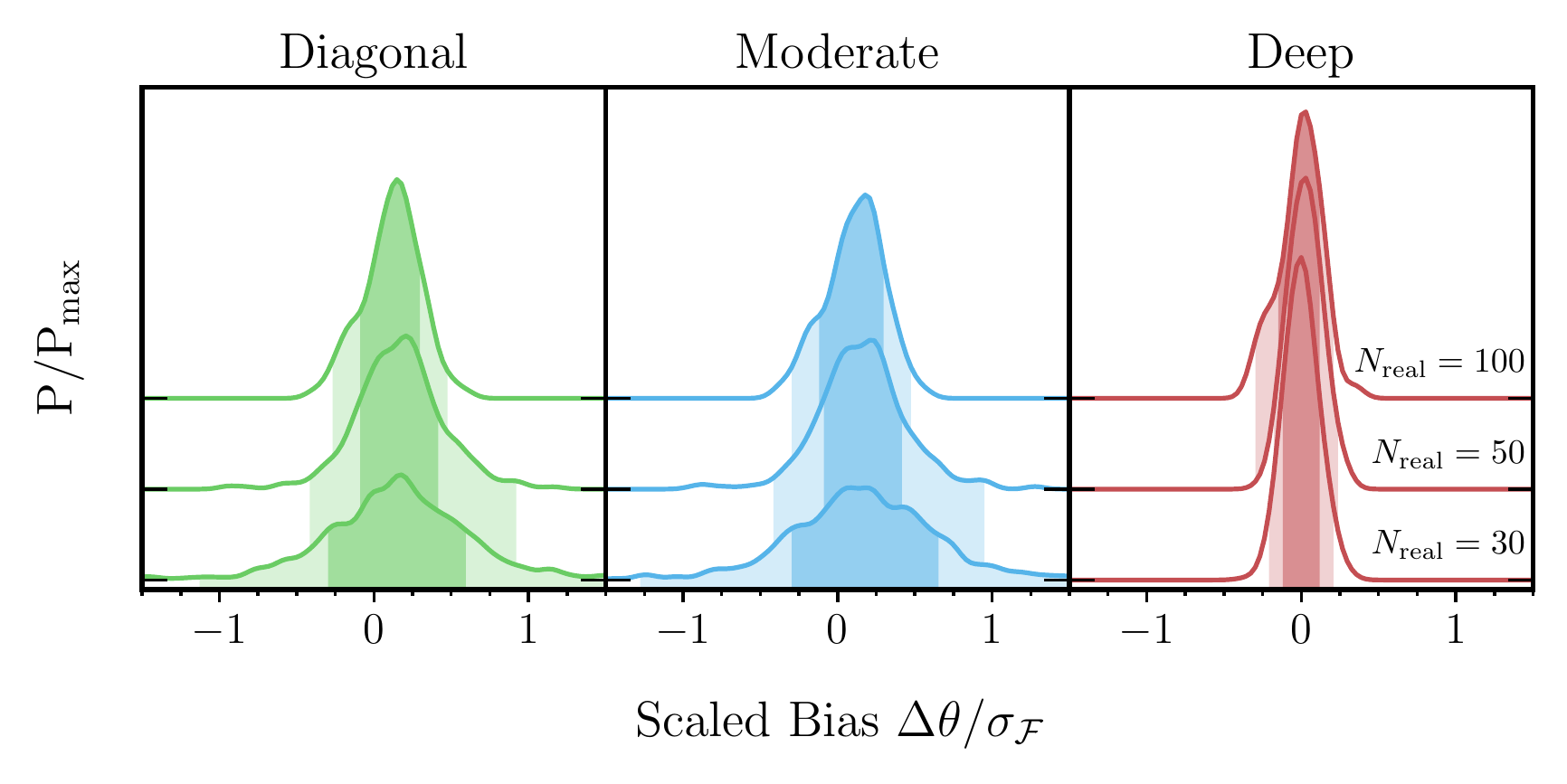}
\centering
\caption{ \label{fig:bias}
Noise in the covariance matrix leads to additional scatter in the parameter constraints.
We show the probability density function (PDF) of the ratio of the parameter bias to the fiducial uncertainty, $\Delta\theta/\sigma_\mathcal{F}$, for $\sigma_{\rm map} = 10\,\mu$K-arcmin simulations for diagonal (green), moderate (blue), and deep conditioning (red).
With increasing vertical offset we show the $\Nreal=30, 50$, and 100 simulations and indicate the 68\% and 95\% interval in faint colour under the curves.
Diagonal and moderate conditioning lead to a substantial scatter in the recovered parameters, as shown by the broad 68\% and 95\% intervals.
This extra noise is not accounted for in the data likelihood and independent of the change to the curvature of the likelihood.
Increasing $\Nreal$ reduces uncertainty on the covariance estimate and leads to narrower distributions, i.e. the scatter to parameter constraints is decreased.
Deep conditioning performs almost as well as the fiducial covariance matrix, even for $\Nreal = 30$.
Moreover, with increasing $\Nreal$ the distributions approach a Gaussian, with KS-test p-values in the 95\% interval for all conditioning schemes given $\Nreal=100$.
The PDFs shown here are estimated by computing a histogram of the scaled bias and smoothing it using a Gaussian kernel.
}
\end{figure*}

By considering the number of noise-variance dominated bins estimated, we can understand how these results change for: different map-noise levels; different band power bin widths; and subsets of the \TT{}, \TE, and \EE{} power spectra.
Lowering the map-noise level increases the angular multipole range over which the well-measured sample-variance part of the covariance dominates.
This boosts the performance of the conditioning schemes; for example, for $\Nreal=30$, $\sigma_{\rm map} = 2\,\mu$K-arcmin simulations, the uncertainty underestimation of moderate conditioning is reduced from $3.5^{+2.7}_{-2.3}\%$ to $0.5^{+1.4}_{-1.2}\%$.
Similarly, when doubling the bin size from $\Delta\ell=50$ to $\Delta\ell=100$, moderate conditioning recovers the correct parameter uncertainty to $0.1^{+1.2}_{-1.3}\%$ (at $\sigma_{\rm map} = 10\,\mu$K-arcmin).
Since the temperature power spectrum is sample-variance dominated over the largest angular multipole range, we observe more robust constraints from \TT{}-only data: moderate conditioning yields the correct parameter uncertainty at $1.3^{+2.7}_{-2.5}\%$.
The improvement in each case is owed to the reduction in the number of covariance elements estimated from the limited data realisations.

\begin{figure}
\includegraphics[width=3.464in]{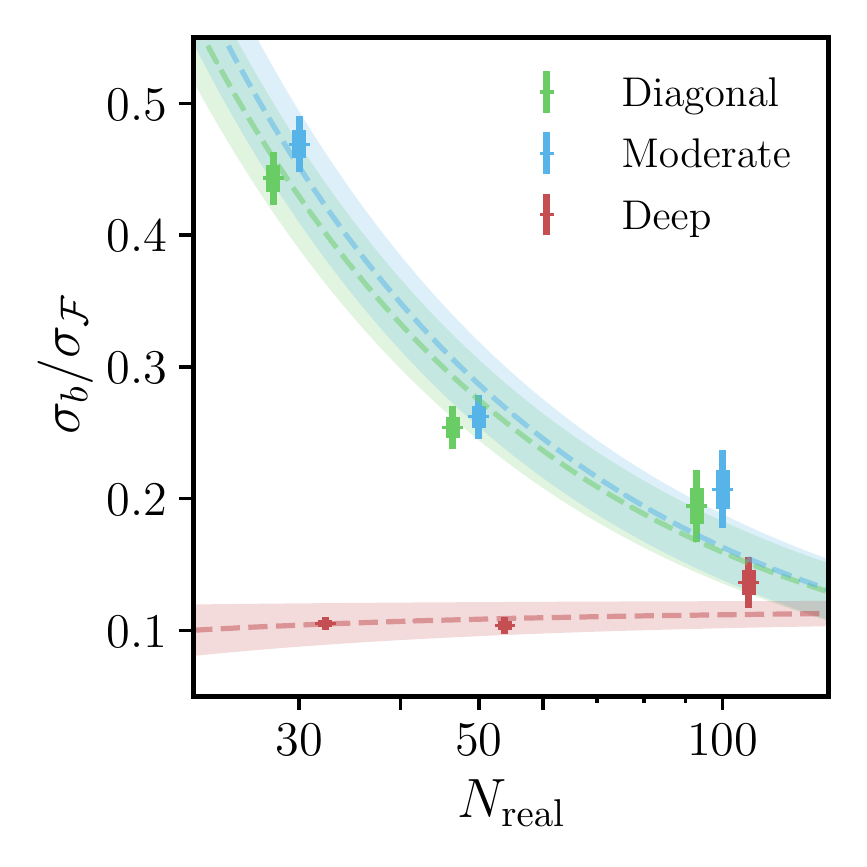}
\centering
\caption{ \label{fig:bias_nbundle}
Errors on the covariance matrix introduce additional (and unaccounted for) scatter in the recovered parameters.
We show here the ratio of the additional scatter to the fiducial parameter uncertainty for simulations with $\sigma_{\rm map} = 10\,\mu$K-arcmin, parametrised by $\sigma_b$, the half-width of the 68\% interval of $\theta/\sigma_\mathcal{F}$.
The wide and thin markers indicate the $1-2\sigma$ range based on the number of samples for diagonal (green), moderate (blue), and deep conditioning (red), respectively.
The dotted lines show an inverse power-law fit with slope $\Nreal^{-1}$ and the semi-transparent region corresponds to the $1\sigma$ region around it.
Due to the small dataset, these fits only serve to guide the eye and are not statistically meaningful.
The additional scatter to parameter constraints is reduced as the covariance estimate improves with higher $\Nreal$.
We do not expect to be able to reach zero extra noise due to residual noise in the fiducial covariance matrix.
}
\end{figure}

\subsubsection{Bias}
\label{subsubsec:bias}

We next turn to evaluating the level of bias in the recovered parameter constraints due to the noisy covariance estimate.
We define the bias $\Delta \theta$ as the median difference of the best-fit values of a given conditioning scheme compared to the best-fit value of the fiducial covariance matrix. 
We also measure the additional scatter (or noise) introduced in the best-fit values by the noisy covariance.
We quote both quantities relative to the expected uncertainty, by dividing the measured bias or scatter by the expected uncertainty from a Fisher forecast, $\sigma_{\mathcal{F}}$.\footnote{We use the Fisher forecast to avoid introducing additional noise from the individual error bar estimates, although we find similar results in both cases.}

We use the median and $\sigma_b$, defined as half the central 68\% interval, to describe the two key quantities of interest, the bias and the extra noise to the best-fit point, respectively.
The bias is a global shift to the parameter constraint and depends on the specific point in parameter space being explored.
In theory, this can be corrected for using a set of simulations.
The extra noise on the other hand can be understood as a random shift to the best-fit value, which is unique for each individual realisation.
The additional scatter is not captured by the data likelihood and independent of the change to the curvature of the likelihood explored in \S\ref{subsubsec:unc}.
This effect needs to be accounted for explicitly through a modification of the likelihood or similar procedure.

Scaled bias distributions are shown in Figure \ref{fig:bias} for the $\sigma_{\rm map} = 10\,\mu$K-arcmin simulations.
For $\Nreal = 30$, diagonal and moderate conditioning lead to a sizeable scatter of parameter constraints, with $\sigma_b / \sigma_{\mathcal{F}} = 0.44$ and $0.47$, respectively.
There is little difference in the best-fit values obtained from these two schemes, which indicates that this extra noise is driven by the uncertainty on the main diagonal elements of the covariance matrix estimates.
With medians of $0.16$ and $0.17$ for diagonal and moderate conditioning, respectively, we detect a bias at $\sim$10$\,\sigma$.
However, this bias is drowned out by the random scatter to the best-fit point.
Deep conditioning does not admit a sizeable bias or significant extra noise.
The median of the distribution is consistent with zero and $\sigma_b / \sigma_{\mathcal{F}} = 0.11$.

We do not expect to reach zero scatter to the best-fit point, due to residual noise in the fiducial covariance.
We quantify this lower limit by running an additional set of simulations with the same setting and benchmarking the performance of the associated new fiducial covariance using the original simulations.
We find that the half-width of the 68\% interval of the scaled bias distribution spans $0.07$.
This is over half of the level of scatter observed for deep conditioning quoted above, indicating that the random shifts admitted by the conditioning scheme are in practice appreciably lower and typically below the $0.1 \sigma_{\mathcal{F}}$ level.

Figure \ref{fig:bias} also allows us to observe the evolution of the distributions with $\Nreal$.
As expected, the distributions become tighter as more realisations become available, i.e. the 68\% and 95\% intervals shrink, and the median values move closer towards zero, as the covariance estimate is improved.
Practically, this means that the scatter to the parameter constraints, $\sigma_b$, shrinks and the uncertainty approaches the fiducial value.
We focus on the evolution of the extra noise with $\Nreal$ in Figure \ref{fig:bias_nbundle}, which shows this trend more clearly.
However, even at $\Nreal=100$, diagonal and moderate conditioning both still lead to non-negligible scatter of the best-fit parameters, with $\sigma_b / \sigma_{\mathcal{F}} = 0.19$ and $0.21$, respectively.
We detect no significant change in the performance of deep conditioning with $\Nreal$.

The extra noise dominates over the change to the curvature of the likelihood seen in \S\ref{subsubsec:unc} for most simulations tested.
We combine the two effects by adding the standard deviation of the bias distributions to the parameter uncertainties measured in \S\ref{subsubsec:unc} in quadrature.
Generally, the total effect is an increase in the parameter uncertainty compared to the fiducial case.
For the $\Nreal=30$, $\sigma_{\rm map} = 10\,\mu$K-arcmin simulations, we find the uncertainty on the recovered best fit parameter to be a factor of 1.3 larger than the apparent posterior width from the likelihood for diagonal and moderate conditioning.
This corresponds to an increase by a factor of 1.2 and 1.3 compared to when the true covariance is used, respectively.
Diagonal conditioning leads to a total uncertainty closer to unity, because the scatter of the best-fit point is offset by the the larger uncertainty underestimation from the curvature of the likelihood.
This balance changes with increasing $\Nreal$, as the extra noise shrinks but the likelihood curve obtained from diagonal conditioning remains narrower than the fiducial case, leading to a total underestimation of the parameter uncertainty by $2.4^{+1.5}_{-1.2}\%$.
For $\Nreal \geq 100$ the latter is no longer an issue for moderate conditioning, and as the scatter of the best-fit point is reduced the total uncertainty converges to the fiducial value.
Deep conditioning recovers the fiducial uncertainty to less than $2\%$ in all cases tested.

We expect that the effect of noise in the covariance on the parameter constraints will scale as the ratio of the number of bins estimated to the number of realisations \citep{dodelson13}.
For a fixed number of realisations, reducing the number of bins estimated will also reduce the additional uncertainty incurred due to the covariance estimate for all conditioning schemes.
We have demonstrated this on all three axes: changing the map-noise level (varying the number of band powers dominated by noise variance); changing the number of band power bins; and reducing the number of spectra from \TT{}, \TE{} and \EE{} to \TT{}-only.
For instance, for $\Nreal=30$ and moderate conditioning, lowering the noise from $\sigma_{\rm map} = 10\,\mu$K-arcmin to $\sigma_{\rm map} = 2\,\mu$K-arcmin reduces $\sigma_b / \sigma_{\mathcal{F}}$ from $0.47$ to $0.17$.
Similarly, halving the number of bins through a straightforward rebinning from $\Delta\ell=50$ to $\Delta\ell=100$ decreases the random scatter to $0.21$ for moderate conditioning (at $\sigma_{\rm map} = 10\,\mu$K-arcmin).
Lastly, analysing only the temperature power spectrum reduces $\sigma_b / \sigma_{\mathcal{F}}$ to $0.16$, since the \TT{} band powers are sample-variance dominated over the largest angular multipole range.

While the tests so far have focused on single-parameter fits, we are generally interested in the size of the additional scatter due to noise in the covariance matrix in multiple-parameter fits, for instance to the full $\Lambda\mathrm{CDM}$ model.
We use the best-fit points of a subset of 100 $\Nreal = 30$, $\sigma_{\rm map} = 10\,\mu$K-arcmin simulations and look at the additional scatter scaled by the fiducial parameter covariance matrix described in \S\ref{subsec:parameter_comparison}.
We find that all cosmological parameters show a similar level of additional scatter in the marginalised 1D posteriors.
Taking the geometric mean across the five dimensions, the extra scatter with diagonal, moderate, and deep conditioning is $\sigma_b / \sigma_0 = 0.44$, $0.50$ and $0.11$ respectively, statistically consistent with the results of $0.44$, $0.47$ and $0.11$ in the single-parameter fits.
Note, however, that this scatter applies to each individual axis in the parameter space, so if one looks at the distribution of the distance between the true and recovered best-fit across $N$ dimensions, the width expands by up to a factor of $\sqrt{N}$.
We conclude that the estimates of the additional scatter due to noise in the covariance estimate in \S\ref{subsec:parameter_comparison} and Fig.~\ref{fig:bias_nbundle} are applicable to the marginalised 1D posterior constraints of each parameter in the full $\Lambda\mathrm{CDM}$ model.

\section{Conclusion}
\label{sec:conclusion}


In this work, we assess how effective conditioning noisy estimates of the covariance matrix is at producing robust parameter constraints. 
We benchmark the performance of four conditioning schemes imposing increasingly strong priors on the estimated covariance matrix: no conditioning, diagonal conditioning, moderate conditioning, and deep conditioning.
We simulate \TT{}+\TE{}+\EE{} power spectrum analyses spanning an angular multipole range of $300 \leq \ell < 3500$ with band powers of width $\Delta\ell = 50$, mimicking access to $\Nreal = 30, 50,$ or 100 data realisations and map-noise levels of $\sigma_{\rm map} = 10, 6.4$ and $2\,\mu$K-arcmin.
We find that, for all cases tested, conditioning the covariance matrix is necessary to obtain robust parameter constraints. 

When examining the resulting parameter constraints, the largest effect of noise in the covariance estimate is a significant scatter in the best-fit points.
Crucially, this additional uncertainty is not reflected in the reported parameter posteriors, from e.g., an MCMC analysis of the data likelihood, and must be explicitly accounted for by some other means. 
Not accounting for this additional scatter can significantly underestimate the uncertainty on cosmological parameters.
For $\Nreal = 30$, $\sigma_{\rm map} = 10\,\mu$K-arcmin simulations, we find that the parameter uncertainties are a factor of 1.3 larger than the apparent posterior width from the likelihood for diagonal and moderate conditioning.
Importantly, the distribution of this additional scatter is significantly non-Gaussian. 
The stricter priors of deep conditioning eliminate the increase in uncertainty.
Increasing the number of data-splits used in the covariance estimate also improves the situation: with 100 data realisations, we report no increase in the parameter uncertainty for diagonal and moderate conditioning.
In many cases, we expect it will be necessary to add an explicit correction for this additional scatter in order to accurately capture the parameter uncertainty.
While the optimal approach may vary between analyses, we alert the reader to \citet{sellentin16} and \citet{percival21}, who offer solutions in the Bayesian and Frequentist inference frameworks.

We also report two other less significant phenomena: a distortion of the curvature of the likelihood and a bias of parameter constraints.
For substantial noise in the covariance estimate ($\Nreal = 30$, $\sigma_{\rm map} = 10\,\mu$K-arcmin), diagonal and moderate conditioning schemes lead to an artificially tight likelihood; this translates into an underestimation of the parameter uncertainty by a few percent.
More data realisations ameliorates this underestimate. 
However, for diagonal conditioning, which does not capture the full correlation structure of the data, the likelihood does not approach the fiducial curve and parameter errors remain modestly too small by $2.4^{+1.5}_{-1.2}\%$ at $\Nreal = 100$.
While we detect a small bias to parameter constraints induced by the residual noise in the covariance matrix, it is smaller than the additional scatter in all cases.
Both the bias and uncertainty underestimation approach zero as $\Nreal$ increases and the covariance estimate improves; the additional scatter is more significant than both effects at all times. 

At a fixed number of data realisations, we expect the covariance matrix to become better constrained as the number of band powers decreases as there are fewer matrix elements (with the same fractional error per element). 
In the tests in this work, where the sample variance contributions were estimated from a much larger number of simulations, this means the effects of uncertainty in the covariance matrix should fall at lower map-noise levels as well as when looking at a brightest selection of \TT{}, \TE{} and \EE{} spectra, or widening the band power angular multipole bins. 
We confirm that this intuition is correct and find that these changes indeed lead to a reduction in the additional scatter, misestimation of the likelihood curvature, and bias.

We are interested in expanding upon this work, by exploring parameter constraints for a wider and finer grid of analysis parameters, including $\Nreal$, $\sigma_{\rm map}$, the angular multipole range, the band power bin width, the number of transfer-function simulations, inclusion of the lensing power spectrum, and increasing the sky area observed.
This extended set of simulations would allow us to provide scaling relations for the parameter uncertainty misestimation and bias for different conditioning schemes as these analysis parameters are varied.
This would ultimately allow us to produce a practical guide for readers to determine what amount of conditioning is necessary for a given analysis.
Additionally, we would like to explore further how parameter constraints behave in a multi-dimensional model space.
Using interpolation approaches to obtain model CMB power spectra \citep[e.g.,][]{fendt07a, alessio2021} may make full explorations of the parameter posteriors computationally tractable.
Current-generation ground-based CMB experiments have the potential of making exciting discoveries of physics beyond the standard model.
Understanding and managing the sources of even small changes to parameter uncertainties and biases is crucial to producing reliable results.

\section*{Acknowledgments}
The authors thank Piaera Lauritz for helpful advice on the nomenclature of conditioning schemes.
We acknowledge support from the University of Melbourne and an Australian Research Council Future Fellowship (FT150100074).
This research used resources of the National Energy Research Scientific Computing Center (NERSC), a U.S. Department of Energy Office of Science User Facility operated under Contract No. DE-AC02-05CH11231.
The data analysis presented uses the scientific python stack \citep{hunter07, jones01, vanDerWalt11}.

\section*{Data Availability}

The data underlying this article will be shared on reasonable request to the corresponding author.

\begin{appendices}
\section*{Appendix}
\label{app}
We present the uncertainty underestimation and shift to the best-fit point from joint \TT{}+\TE{}+\EE{} constraints in Table \ref{tab:unc_bias}. The results cover diagonal, moderate, and deep conditioning for $\Nreal = 30, 50, 100$, and $\sigma_{\rm map} = (10, 6.4, 2)\,\mu$K-arcmin.
In Table \ref{tab:fields}, we show the same statistics for all combinations of the \TT{}, \TE{}, \EE{} spectra of the $\Nreal = 30$, $\sigma_{\rm map} = 10\,\mu$K-arcmin simulations.

\clearpage

\begin{table*}
\def\arraystretch{2.0}
\footnotesize
\setlength{\tabcolsep}{3pt}
\centering
\caption[
Uncertainty increase
]{
Median and 68\% interval of the uncertainty underestimation in percent compared to the performance of the fiducial matrix and shift to the best-fit point of $\theta_{MC}$ constraints scaled by the predicted uncertainty from a Fisher forecast for simulations with $\Nreal=30, 50, 100$ and $\sigma_{\rm map} = (10, 6.4, 2)\,\mathrm{\mu K - arcmin}$ for diagonal, moderate, and deep covariance conditioning.
}
\begin{tabular}{c c c c c c c c c c c c c }
\hline
\multirowcell{3}{\vspace{1cm}\\Noise -- level\\ $\left[ \mathrm{\mu - arcmin} \right] $}
& \multicolumn{12}{c}{$N_{\rm real}$}\\
\cmidrule{3-13}
&
& \multicolumn{3}{c}{30}
&
& \multicolumn{3}{c}{50}
&
& \multicolumn{3}{c}{100} \\
\cmidrule{3-5} \cmidrule{7-9} \cmidrule{11-13} 
&
& \rotatebox[origin=c]{60}{Diagonal}
& \rotatebox[origin=c]{60}{Moderate}
& \rotatebox[origin=c]{60}{Deep}
& \rotatebox[origin=c]{60}{\phantom{xxxxxxxxx}}
& \rotatebox[origin=c]{60}{Diagonal}
& \rotatebox[origin=c]{60}{Moderate}
& \rotatebox[origin=c]{60}{Deep}
& \rotatebox[origin=c]{60}{\phantom{xxxxxxxxx}}
& \rotatebox[origin=c]{60}{Diagonal}
& \rotatebox[origin=c]{60}{Moderate}
& \rotatebox[origin=c]{60}{Deep}\\
& \vspace{-0.5cm} \\
\hline \hline
\multicolumn{13}{l}{Uncertainty Underestimation: $1-\sigma/\sigma_0\,[\%]$}\\
10 &
&
$5.8^{+2.1}_{-1.9}$ &
$3.5^{+2.7}_{-2.3}$ &
$0.3^{+1.5}_{-1.5}$ &
&
$4.4^{+1.8}_{-1.6}$ &
$2.1^{+2.1}_{-1.9}$ &
$0.2^{+1.6}_{-1.6}$ &
&
$2.4^{+1.5}_{-1.2}$ &
$0.0^{+1.5}_{-1.4}$ &
$-0.7^{+1.2}_{-1.4}$ \\
6 &
&
$4.5^{+1.8}_{-1.6}$ &
$2.1^{+2.1}_{-1.9}$ &
$0.0^{+1.3}_{-1.3}$ &
&
$3.5^{+1.4}_{-1.2}$ &
$0.9^{+1.6}_{-1.4}$ &
$-0.2^{+1.3}_{-1.3}$ &
&
$3.4^{+1.0}_{-1.3}$ &
$0.6^{+1.5}_{-1.3}$ &
$-0.2^{+1.5}_{-0.9}$ \\
2 &
&
$3.3^{+1.1}_{-1.0}$ &
$0.5^{+1.4}_{-1.2}$ &
$-0.3^{+1.1}_{-1.1}$ &
&
$3.3^{+0.9}_{-1.0}$ &
$0.5^{+1.1}_{-1.1}$ &
$-0.1^{+1.1}_{-1.1}$ &
&
$2.8^{+0.9}_{-1.1}$ &
$0.1^{+1.2}_{-1.3}$ &
$-0.4^{+1.0}_{-1.2}$ \\
\hline
\multicolumn{13}{l}{Best-fit Shift: $\Delta\theta/\sigma_{\mathcal{F}}$}\\
10 &
&
$0.16^{+0.45}_{-0.44}$ &
$0.17^{+0.46}_{-0.47}$ &
$0.00^{+0.10}_{-0.11}$ &
&
$0.17^{+0.25}_{-0.26}$ &
$0.16^{+0.25}_{-0.27}$ &
$0.03^{+0.11}_{-0.10}$ &
&
$0.13^{+0.22}_{-0.17}$ &
$0.13^{+0.24}_{-0.17}$ &
$0.00^{+0.14}_{-0.13}$ \\
6 &
&
$0.14^{+0.31}_{-0.31}$ &
$0.13^{+0.31}_{-0.35}$ &
$0.02^{+0.11}_{-0.11}$ &
&
$0.12^{+0.24}_{-0.21}$ &
$0.11^{+0.24}_{-0.20}$ &
$0.01^{+0.11}_{-0.12}$ &
&
$0.07^{+0.15}_{-0.15}$ &
$0.07^{+0.16}_{-0.15}$ &
$-0.01^{+0.12}_{-0.13}$ \\
2 &
&
$0.09^{+0.16}_{-0.18}$ &
$0.08^{+0.16}_{-0.18}$ &
$0.02^{+0.13}_{-0.12}$ &
&
$0.08^{+0.14}_{-0.12}$ &
$0.08^{+0.14}_{-0.13}$ &
$0.02^{+0.12}_{-0.12}$ &
&
$0.11^{+0.14}_{-0.14}$ &
$0.11^{+0.15}_{-0.15}$ &
$0.04^{+0.17}_{-0.15}$ \\
\hline
\end{tabular}
\label{tab:unc_bias}
\end{table*}

\begin{table*}
\def\arraystretch{2.0}
\footnotesize
\setlength{\tabcolsep}{3pt}
\centering
\caption[
Fields
]{
Median and 68\% interval of the uncertainty underestimation in percent compared to the fiducial matrix and shift to the best-fit point of $\theta_{MC}$ constraints scaled by the predicted uncertainty from a Fisher forecast for the $\Nreal=30, \sigma_{\rm map} = 10\mathrm{\mu K - arcmin}$ simulations for diagonal, moderate, and deep covariance conditioning for any combination of the \TT{}, \TE{}, and \EE{} power spectra.
}
\begin{tabular}{c c c c c c c c c}
\hline
\multirowcell{2}{\thead{\\Spectra used\\in Likelihood}}
&
& \multicolumn{3}{c}{\thead{Uncertainty\\Underestimation:\\$1-\sigma/\sigma_0\,[\%]$}}
&
& \multicolumn{3}{c}{\thead{Best-fit Shit:\\$\Delta\theta/\sigma_{\mathcal{F}}$}}\\
\cmidrule{3-5}
\cmidrule{7-9}
& \rotatebox[origin=c]{60}{\phantom{xxxxxxxxx}}
& \rotatebox[origin=c]{60}{Diagonal}
& \rotatebox[origin=c]{60}{Moderate}
& \rotatebox[origin=c]{60}{Deep}
& \rotatebox[origin=c]{60}{\phantom{xxxxxxxxx}}
& \rotatebox[origin=c]{60}{Diagonal}
& \rotatebox[origin=c]{60}{Moderate}
& \rotatebox[origin=c]{60}{Deep}\\
\hline \hline
TT, EE, TE &
&
$5.8^{+2.1}_{-1.9}$ &
$3.5^{+2.7}_{-2.3}$ &
$0.3^{+1.5}_{-1.5}$ &
&
$0.16^{+0.45}_{-0.44}$ &
$0.17^{+0.46}_{-0.47}$ &
$0.00^{+0.10}_{-0.11}$ \\
TT &
&
$4.8^{+2.2}_{-2.1}$ &
$1.3^{+2.7}_{-2.5}$ &
$0.6^{+2.0}_{-2.1}$ &
&
$0.07^{+0.15}_{-0.15}$ &
$0.07^{+0.15}_{-0.16}$ &
$0.00^{+0.07}_{-0.07}$ \\
TE &
&
$2.8^{+1.9}_{-2.0}$ &
$1.4^{+2.4}_{-2.5}$ &
$0.4^{+2.1}_{-2.0}$ &
&
$0.02^{+0.15}_{-0.15}$ &
$0.01^{+0.15}_{-0.15}$ &
$0.00^{+0.06}_{-0.07}$ \\
EE &
&
$3.9^{+3.3}_{-3.1}$ &
$3.2^{+4.6}_{-4.3}$ &
$0.4^{+2.4}_{-2.5}$ &
&
$0.10^{+0.27}_{-0.30}$ &
$0.10^{+0.30}_{-0.34}$ &
$0.00^{+0.07}_{-0.08}$ \\
TT, TE &
&
$4.7^{+1.9}_{-1.8}$ &
$2.1^{+2.1}_{-2.1}$ &
$0.4^{+1.8}_{-1.6}$ &
&
$0.10^{+0.23}_{-0.23}$ &
$0.09^{+0.23}_{-0.23}$ &
$0.00^{+0.09}_{-0.08}$ \\
TT, EE &
&
$5.4^{+2.4}_{-2.3}$ &
$3.1^{+3.0}_{-2.7}$ &
$0.5^{+1.7}_{-1.8}$ &
&
$0.10^{+0.31}_{-0.41}$ &
$0.13^{+0.36}_{-0.43}$ &
$0.00^{+0.08}_{-0.09}$ \\
EE, TE &
&
$4.1^{+2.4}_{-2.2}$ &
$2.9^{+3.1}_{-2.8}$ &
$0.3^{+1.8}_{-1.8}$ &
&
$0.09^{+0.31}_{-0.32}$ &
$0.09^{+0.32}_{-0.35}$ &
$0.02^{+0.10}_{-0.09}$ \\
\hline
\end{tabular}
\label{tab:fields}
\end{table*}
\end{appendices}

\clearpage

\bibliography{spt}

\end{document}